\documentclass[twocolumn,showpacs,preprintnumbers,amsmath,amssymb,prb]{revtex4-1}
\usepackage{soul}
\usepackage[bbgreekl]{mathbbol}
\usepackage{graphicx}
\usepackage{dcolumn}
\usepackage{bm}
\usepackage{mathtools}
\usepackage{hyperref}
\usepackage{color}
\graphicspath{{fig/}{img/}}
\DeclareMathAlphabet\mathbfcal{OMS}{cmsy}{b}{n}

\renewcommand{\l}{{\b}}

\renewcommand{\a}{{\bf a}}

\renewcommand{\b}{{\bf b}}
\newcommand{\vd}{v}
\newcommand{\kn}{{\bf  k}}
\newcommand{\qn}{{\bf  q}}
\newcommand{\Kn}{{\bf  K}}
\newcommand{\Rn}{{\bf R}}

\newcommand{\rn}{{\bf r}}

\def\gsim{\lower.35em\hbox{$\stackrel{\textstyle>}{\textstyle\sim}$}}
\def\lsim{\lower.35em\hbox{$\stackrel{\textstyle<}{\textstyle\sim}$}}

\begin{document}
\title{Quantum Internal Structure of Plasmons}
\author{Jinlyu Cao}
\affiliation{  Department of Physics, Indiana University, Bloomington, IN 47405 and \\
Quantum Science and Engineering Center, Indiana University, Bloomington, IN, 47408}
\author{H.A. Fertig}
\affiliation{  Department of Physics, Indiana University, Bloomington, IN 47405 and \\
Quantum Science and Engineering Center, Indiana University, Bloomington, IN, 47408}
\author{Luis Brey}
\affiliation{Materials Science Factory, Instituto de Ciencia de Materiales de Madrid (CSIC), Cantoblanco, 28049 Madrid, Spain}

\date{\today}
\begin{abstract}
Plasmons are usually described in terms of macroscopic quantities such as electric fields and currents.   However as fundamental excitations of metals they are also quantum objects with internal structure. We demonstrate that this can induce an intrinsic dipole moment which is tied to the quantum geometry of the Hilbert space of plasmon states.  This {\it quantum geometric dipole} offers a unique handle for manipulation of plasmon dynamics, via density modulations and electric fields.  As a concrete example we demonstrate that scattering of plasmons with non-vanishing quantum geometric dipole from impurities is non-reciprocal, skewing in different directions in a valley-dependent fashion.
This internal structure can be used to control plasmon trajectories in two dimensional materials.
\end{abstract}
\maketitle

\noindent {\it Introduction.}
Plasmons are fundamental excitations of metals in which interactions lock electrons into coherent oscillatory motion.  In recent years controlling their dynamics has become increasingly important as applications in information processing and communication have been realized \cite{Hutter_2004,Pitarke:2006aa,Sekhon_2011,Nikitin:2011aa,Grigorenko:2012aa}.  Moreover, advances in two-dimensional material fabrication have allowed great strides in realizing new platforms for plasmons \cite{Wunsch_2006,Hwang_2007,Thygesen_2017,Agarwal_2018}, where strong coupling between electromagnetic waves and electrons \cite{Linic_2011,Ju_2011,Stauber_2020}, low loss energy propagation \cite{Woessner_2015,Alcaraz_2018,Giri_2020}, and fundamentally new types of plasmon dispersions \cite{Ni_2015,Brey:2020aa} may all be realized.

Because of their collective nature plasmons are usually described in terms of macroscopic quantities, typically electric fields and currents \cite{Pitarke:2006aa,Grigorenko:2012aa,Nikitin:2011aa,Kumar:2016aa,Stauber_2020}.   As basic excitations of metals, however, they are also quantum bosonic quasiparticles which may carry internal, microscopic structure.  Such structure offers new avenues for control and interrogation of plasmons, allowing windows on their fundamental properties which are otherwise difficult to access.  In this work, we demonstrate that under appropriate circumstances such structure {\it must} be present, due to the quantum geometry of the plasmon Hilbert space.  This takes the form of a dipole moment directly tied to the plasmon momentum, which can be properly understood as a {\it quantum geometric dipole} (QGD) \cite{Cao_2021}.  Its existence suggests new ways of manipulating plasmons: for example, a density step in a two-dimensional metal should bind plasmons, moving them in opposite directions depending on the valley in which they reside, offering a way to incorporate plasmons into valleytronic systems.

An important consequence of the plasmon QGD, which we analyze in detail in what follows, is that it leads to non-reciprocal scattering \cite{Nagaosa:2010aa,Sinitsyn:2007aa,Glazov_2020} of plasmons from impurities.  Indeed, within a microscopic RPA treatment of plasmon wavefunctions, we show that asymmetry in scattering around the forward direction is directly proportional to the QGD itself.  We also develop an effective macroscopic description of the scattering, where non-reciprocity is evident in situations for which plasmons should present non-vanishing QGD's.  While macroscopic descriptions also predict non-reciprocal behavior in boundary reflection \cite{Shi_2018,SI_Shi_discussion},
observation of plasmon skew scattering from, for example, Coulomb impurities, affords direct confirmation that the plasmons carry a microscopic internal dipole moment.  Thus, to our knowledge, observation of this physics -- as should be possible with near-field microscopy --
would offer the first demonstration that the internal {\it quantum} structure of plasmons can play a direct and important role in their dynamics.

 \begin{figure}[tbp]
\includegraphics[width=9.0cm,clip]{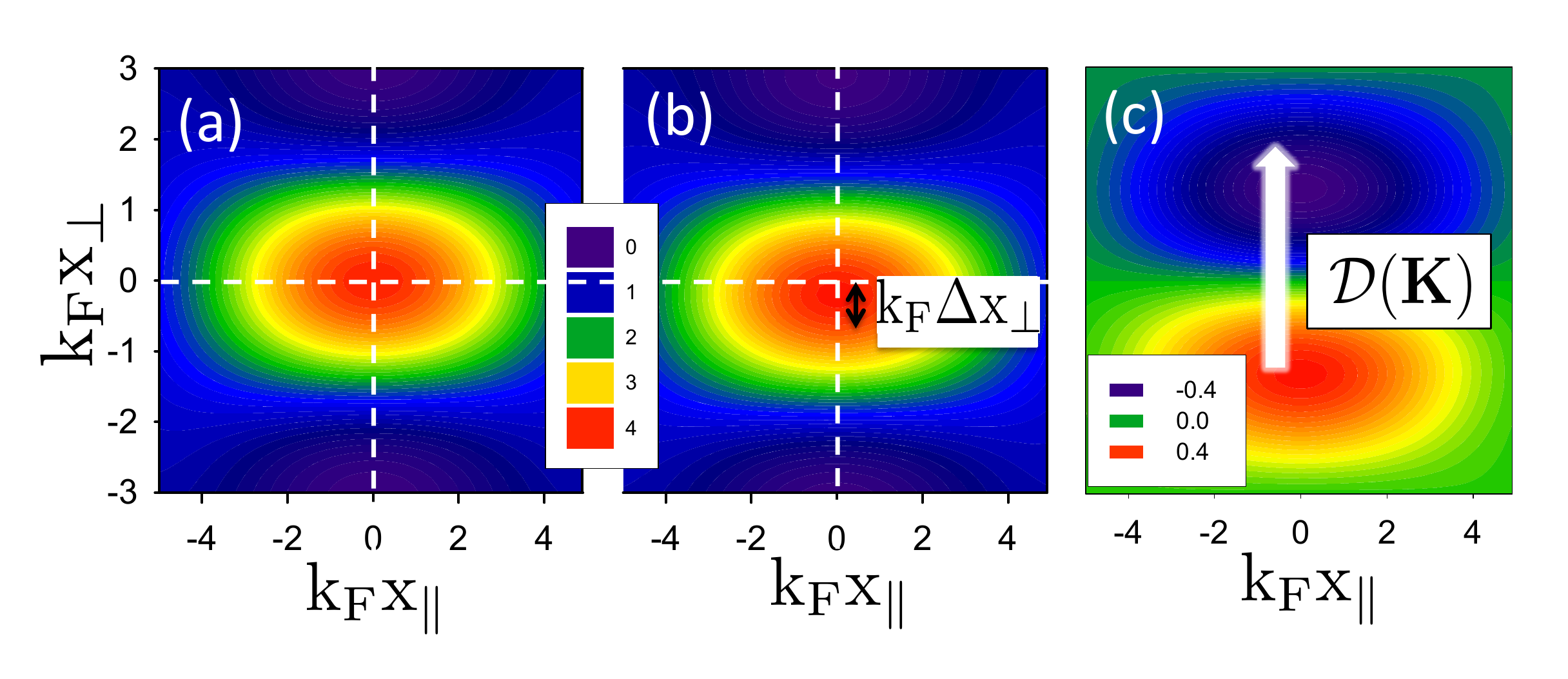}
\caption{
Square of the plasmon wavefunction, $| \Phi _{\Kn} | ^2$,  for $\alpha K k_F \equiv \frac {\hbar ^2 v^2 }{4\delta ^2}K k_F =0$ (a) and $=0.1$ (b), vs. electron-hole relative position.
$x_{\perp}$ and $x_{\parallel}$  are the spatial coordinates perpendicular and parallel to the plasmon wavevector respectively.  The finite  $\Kn$ breaks rotational symmetry and the wavefunction has an ellipsoidal shape which is more extended in the direction parallel to $\Kn$.
For finite $\alpha$ (b), the wave function is not centered at
$x_{\perp}=0$ and a quantum geometrical dipole in this direction appears. This is evident in (c), the difference between the square of the  wavefunctions (a) and (b).
}
\label{Figure1}
\end{figure}

\vspace{0.1cm}
\noindent {\it Hamiltonian and Plasmon Wavefunction.}
We begin by adopting a simple model for electrons described by a gapped two-dimensional Dirac Hamiltonian,
\begin{equation}
H_0= \hbar \vd \qn \cdot {\bm \sigma}+ \delta \sigma _z,
\label{H0}
\end{equation}
where $\vd$ is the Dirac velocity, 2$\delta$ is the gap of the system, $\qn$=$(q_x,q_y)$ is the two-dimensional wavevector, and ${\bm \sigma} = (\sigma_x,\sigma_y, \sigma_z)$ are Pauli matrices. This Hamiltonian describes the long wavelength physics {of a single valley} in different materials including gapped graphene \cite{Xiao_2007}, doped transition metal dichalcogenides (TMDs) \cite{Xiao_2012}, and topological insulator surfaces states with a gap opening due to some symmetry-breaking perturbation \cite{Garate_2011,Brey_2014,Reja_2017}.
{(We return to possible impacts of multiple valleys below.)}
For graphene the Pauli matrices act on a sublattice index, while
in the latter two cases the
they act on orbital indices.  For concreteness we consider plasmons in
$n$-doped  systems,
and therefore only include the conduction band of Eq. \ref{H0} with energies
$\epsilon _q = \sqrt{\delta^2 +\hbar ^2 \vd ^2 q^2} $
and wavefunctions $\Psi _{\qn} ({\rn}) \! = \!  \vec {\chi} _{\qn} \frac {e^{i {\qn}\cdot {\rn}}}{\sqrt{S}}$, where $S$ is the system area and
\begin{equation}
 \vec {\chi} _{\qn}
 \! = \!  \left ( \begin{array} {c} e ^{-i \phi_ {\qn}} \sin { \frac {\theta(\qn) }2}  \\
 \cos{\frac {\theta(\qn)} 2 }\end{array} \right ).
 \end{equation}
Here $\theta (\qn)=\tan ^{-1} \frac {\hbar \vd q}{\delta}$ and $\phi _{\qn}=\tan ^{-1} \frac {q_y}{q_x}$.
The conduction band has a Berry's curvature \cite{Berry_1984}
$b _{q}=\frac {\hbar ^2 \vd ^2  \delta}{2 ( \delta ^2 + \hbar ^2 \vd ^2 q ^2) ^{3/2}}$,
which may be taken as $b _q \approx \frac {\hbar ^2 \vd ^2}{ 2 \delta ^2}$ for small $q$.
For TMDs band gaps are of order 1-2$eV$, while plasmon energies are of order $10^{-4}$ eV.  For the concrete examples discussed below, we will assume  $\delta$ to be a large energy scale.
In particular the dispersion may then be approximated by
$\epsilon _q \approx \delta +\frac {\hbar ^2 q^2} {2 m ^* }$ with $m^*=\delta/\vd^2$.

In a quantum description, plasmons are bosonic quasiparticles composed of electron-hole pairs with total momentum ${\bf K}$.  Within the RPA, they can be generated by a quasiparticle creation operator of the form \cite{Sawada:1957aa}
\begin{equation}
Q^{\dag} _{\Kn}=\sum _{\qn} a_{\qn}(\Kn) c^{\dag} _{\qn + {\Kn} } c _{\qn }
\label{wf0}
\end{equation}
where $c _{\qn}$ annihilates an electron with momentum $\qn$, and the coefficients $a_{\qn}(\Kn)$ are complex parameters that need to be determined. The operator  $Q ^{\dag} _{\Kn}$ acts on the Fermi sea with Fermi energy $E_F$ in the conduction band to generate the plasmon state.  Working within RPA (for details see Supplementary Information (SI) \cite{SI}), we obtain explicit plasmon wavefunctions and energies in the form
$\Phi _{\Kn}({\rn,\Rn}) = \frac {e ^{i \Kn \Rn}}S \sum _{\qn}  f_{\qn } (1\! - \! f_{\qn \!+\! {\Kn} })    U(\Kn,\qn)
 e ^{i \qn \cdot \rn}  \,  \, {\rm with}$
\begin{eqnarray}
U(\Kn,\qn) &= &
a_{\qn}(\Kn) \,  \vec {\chi} _{\qn + \Kn}  \otimes \vec {\chi} _{\qn}  ^*
 \, \,  \,  \, \, \, \,  \nonumber \\a_{\qn}(\Kn) &=& \frac {S(\qn+\Kn , \qn )  N(K)}
{\hbar \omega _K \! + \! \epsilon_{\qn } \!  -\!  \epsilon_{\qn +{\Kn} }}, \, \,
\end{eqnarray}
where $\rn=\rn_2 -\rn_1$, $\Rn = \frac {\rn _1+\rn _2} 2$ are relative and center of mass positions for the electron and hole, $f_q$ is the occupation number of the single particle state of momentum $\qn$, and
$S(\qn , \qn ' )= \vec {\chi} ^* _{\qn} \cdot  \vec {\chi} _{\qn '}$.
$N(K)$ is obtained by normalizing the plasmon wavefunction.
The plasmon frequency in the long wavelength limit
has the form
$\omega _K = \sqrt{ 2 \pi \frac {e^2}{\varepsilon _0}  \frac {n_0} { m^*}  K }$ where $n_0$ is the density of electrons in the system.
%
\vspace{0.1cm}
\par \noindent{\it Plasmon Quantum Geometrical Dipole.}
As in the case of single particle states, to examine the quantum geometry associated with plasmon states one must first remove plane wave factors associated with the total momentum $\Kn$. The two-body nature of $\Phi_K$ offers a variety of ways to do this, opening paths to characterize the quantum geometry of its Hilbert space that are inherently multi-body
\cite{Cao_2021}. We define Berry's connections specific to the holes
($j$=1) and electrons ($j$=2), writing
$
{\mathbfcal A } ^{(j)} (\Kn ) = i \langle u_{\Kn,j} | \vec {\nabla} _{\Kn} |u_{\Kn,j} \rangle $
with
$
u_{\Kn,j} = e ^{-i {\Kn} \cdot {\rn}_j} \Phi _{\Kn} ({\rn,\Rn}).
$
These quantities can be directly related to the dipole moment of a plasmon \cite{Cao_2021},
\begin{eqnarray}
{\bf d}& = &  e <\Phi_{\Kn} |\rn _1 - \rn _2 | \Phi _{\Kn}>  \nonumber \\
&=& e\left[{\mathbfcal A}^{(1)}({\bf K})- {\mathbfcal A}^{(2)}({\bf K}) \right] \, \equiv e {\mathbfcal D}(\Kn),
\end{eqnarray}
where ${\mathbfcal D}(\Kn)$ is the quantity we identify as the quantum geometric dipole.  Our formulation shows explicitly that ${\mathbfcal D}(\Kn)$ is determined by the geometry of the Hilbert space of plasmon states \cite{Neuman_2019}, but the connection of the QGD to the electric dipole moment is also evident.
Using the expressions above for wavefunctions of a Dirac fermion, one finds \cite{SI} for long wavelengths,
\begin{equation}\label{eq: QGD aa small K SI}
\mathbfcal{{D}}({\bf K}) =
\frac{2 \alpha }{\left(4 \alpha  k_F^2+1\right)^{3/2}}
    \Big( {\bf K} \times \hat{z} \Big).
\end{equation}
where $\alpha=\hbar^2 v^2/4\delta^2$.
Fig. \ref{Figure1} illustrates the real space form for a plasmon wavefunction for non-vanishing $\alpha$ and for $\alpha=0$.  Note that in the limit of high density, for which the plasmon frequency is high and one does not expect to see quantum effects, $\mathbfcal{{D}}$ vanishes \cite{SI}.
\vspace{0.1cm}
\par \noindent {\it Skew Scattering Due to QGD.}
The intrinsic electric dipole moment of such plasmons suggests they may undergo skew scattered when
impacting upon a charged  impurity.  To see this,
consider  an impurity potential of the form $V_{imp} ({\bf r})=\sum _{\kn} e ^{i \kn \cdot {\bf r}} V^I _k$. Using the fact that the separation of between the electron and hole in the plasmon is small \cite{Egri:1985aa}, we approximate
the potential acting on an electron-hole pair by $V_{imp}({\bf r}_1)-V_{imp}({\bf r}_2) \approx
V^I_{imp} ( {\bf r} _1  - {\rn} _2) \equiv i (\rn _1 -\rn _2) \cdot  \sum _\kn {\bf k} V^I _k e ^{i \kn \cdot \Rn} $.
One may then show that
the transition amplitude between two plasmon states of momenta $\Kn$ and $\Kn '$ of the same frequency is, to lowest order in $(\Kn-\Kn ')$, \begin{eqnarray} & &
< \! \Phi _{\Kn '} | V^I_{imp}| \Phi _{\Kn}\! > = i \left(V^{I} _{|\Kn' \! - \!  \Kn|} \right) {\mathbfcal D}(\Kn) \cdot  (\Kn ' \! \! -\! \!  \Kn)  \, \, .
 \label{scatt}
 \end{eqnarray}
Eq. \ref{scatt} suggests that when a plasmon carries a non-vanishing QGD,
one necessarily finds
non-reciprocal asymmetric skew scattering from an impurity, proportional to $\Kn \times   \Kn '$.
Observation of plasmon skew scattering demonstrates they carry this quantum geometry.


While the above analysis captures the underlying physics of plasmon skew scattering, it fails to capture any non-vanishing forward scattering component that remains when ${\bf K} \leftrightarrow {\bf K}'$, yielding zero in this limit \cite{Efimkin_2012}.  The reason for this is that the effect of the impurity on the ground state density has not been included.  A more complete analysis may be carried out within RPA in which the impurity potential is fully included to first order.  In this approach one computes a {\it correction} to the plasmon operator in Eq. \ref{wf0}, $Q^{\dag}_{{\bf K}_0} = Q^{(0)\dag}_{{\bf K}_0}+ \sum_{\bf K} Q^{(1)\dag}_{{\bf K}_0}({\bf K})$, where $Q^{(1)\dag}$ is linear in the impurity potential.  Viewing $Q^{\dag}_{{\bf K}_0}$ as a bosonic plasmon creation operator allows us to write an effective plasmon Hamiltonian $H=H_0+h_{scat}+h_{scat}^{\dag}$, with $H_0=\sum_{{\bf K}_0} \hbar \Omega({\bf K}_0) Q^{(0)\dag}_{{\bf K}_0}Q^{(0)}_{{\bf K}_0}$ and $h_{scat}=\sum_{{\bf K}_0} \hbar \omega_{{\bf K}_0} Q^{(1)\dag}_{{\bf K}_0} Q^{(0)}_{{\bf K}_0}$, where $\Omega({\bf K}_0)$ includes the linear order correction to the plasmon energy.  Within this approach, the scattering matrix element from a state ${\bf K}_0$ into a state ${\bf K}$ is
\begin{equation}
M({{\bf K},{\bf K}_0}) = \langle 0 | Q^{(0)}_{\bf K} h_{scat} Q^{(0)\dag}_{{\bf K}_0} |0\rangle,
\label{scat_amp}
\end{equation}
where $|0\rangle$ is the vacuum state for plasmons.  The computation of $M$ is lengthy (see SI \cite{SI}); nevertheless, scattering from a Coulomb impurity may be summarized succinctly.  In general $M$ naturally divides into three terms, $M=M^{I}+M^{II}+M^{III}$; in the limit $|{\bf K} - {\bf K}_0| \equiv |\delta {\bf K}| \ll {\bf K_0} \ll k_F$, with $k_F$ the Fermi wavevector, to lowest non-vanishing order in $\delta K$  and to lowest non-trivial order in $\hbar vK_0/\delta$, these terms become
\begin{eqnarray}
M^{I}&=& -i \frac{V_{imp}(\delta K)}{S} {\mathbfcal D}({\bf K}_0) \cdot \delta {\bf K}, \label{fI}\\
M^{II}&=&-\frac{V_{imp}(K) K}{2\pi S k_F}(1 - \frac{\hbar^2 v^2 K^2}{4\delta^2} ) , \label{fII}\\
M^{III}&=&\left[ {{k_FK} \over {2\pi^2}} \left( 1 - \frac{\hbar ^2 v^2 K^2}{4\delta^2} \right) \right] \frac{V(K)}{\hbar\omega_{K}} M^{I}({\bf K},{\bf K}_0),
\label{fIII}
\end{eqnarray}
where $V(K)=2\pi e^2 /\epsilon_0 K$ and $V_{imp}(K)=Z V(K)$ for an impurity of charge $Ze$.  The momentum and angular dependence of these expressions is discussed in the SI \cite{SI}.

These equations may be interpreted as follows.  $M^{I}$ is the direct scattering of a plasmon from the Coulomb impurity, and is equivalent to the Born approximation in Eq. \ref{scatt}.  As presented in Eq. \ref{fI}, $M^I$ is more general than the case of a Coulomb impurity, applying as well to other types of impurity potentials.  Moreover it requires neither specific assumptions about the single-particle electron wavefunctions, nor an assumption of small particle-hole separation.  The contribution of $M^{II}$ may be understood as scattering from the density induced in the ground state by the impurity.  Note this remains non-vanishing in the limit $\delta K \rightarrow 0$, so that this contribution encodes scattering in the forward direction.  This term in principle may also include a skew scattering component, but is of higher order in $\delta K$ than the contributions from $M^I$ and $M^{III}$.  Finally, $M^{III}$ encodes the effect of oscillations in the electric potential induced by the plasmon on its wavefunction, which must be included self-consistently
in RPA.
%
Note this contribution is directly proportional to the QGD; for $\mathbfcal{D}=0$, it makes no contribution.  Interestingly, for long wavelength plasmons (small $K$) this term {\it dominates} over $M^I$, greatly enhancing the skew scattering for small $\delta K$.

While these microscopic descriptions show the crucial role played by the QGD in plasmonic skew scattering, relatively simple results are limited to the near-forward scattering regime.  Moreover obtaining measurable quantities associated with plasmons starting from a microscopic description is rather involved.  As we now discuss, these difficulties can be overcome using a macroscopic analysis of the system.

\vskip 0.1cm

\noindent {\it Macroscopic Plasmon Scattering from Density Inhomogeneities.}
As discussed above, density fluctuations play an important role in both symmetric and skew scattering of plasmons.  In situations where the density varies slowly in space, one can formulate a macroscopic description which captures the full angular dependence of the scattering amplitude.
Suppose the electron density in the two-dimensional metal has the form $n(\rn)$=$ n_0$ +$\delta n (\rn)$, where $\delta n (\rn)$ is the density perturbation induced by a Coulomb impurity.
In this situation plasmons in the system may be analyzed macroscopically by focusing on the dynamics of the electric field in the
metal.  In particular for slowly varying $\delta n (\rn)$ the system may be characterized by a position dependent optical conductivity tensor, $\underline{\sigma }$.
In the local approximation this depends only on the charge density at  $\rn $,
and for a two dimensional metal the diagonal conductivity takes the form $\sigma _{xx}$=$-i \frac {e} {\omega m^*} n(\rn)$.  Crucially, the Hall conductivity is non-vanishing
because of  the Berry's curvature of the bands, and from the Hamiltonian (Eq. \ref{H0})
its form is\cite{Sinitsyn:2006aa}
\begin{equation}
\! \! \!  \sigma _{xy} (\rn) \! \! = \! \! - \sigma _H \! \frac {\delta }{\sqrt{4 \pi \hbar ^2 v^2 n (\rn)/e+\delta ^2}} \! \!
\approx \! \! - \sigma _H \!  + \!   \frac {e \hbar}{2 \delta m ^* } n (\rn)
\label{hall}
\end{equation}
where $\sigma _H$=$\frac {e^2}{4 \pi \hbar}$ is the quantized value of the Hall conductivity when the chemical potential is in the gap.
The right hand term of Eq.\ref{hall} is valid when the Fermi energy measured from the conduction band bottom is much smaller than the gap of the semiconductor.

Consider now a time dependent plasma oscillation described by
an extra charge density modulation,
$ \delta \rho _{\Kn} e^{i ({\Kn \cdot \rn -\omega t)}}$,
with an associated
electric potential
$\phi _{\Kn} e^{i (\Kn \cdot \rn -\omega t)}$, where $\phi _{\Kn}$=$\frac {2 \pi e}{\epsilon _0 K}\delta \rho _{\Kn}$.
Using Ohm's law and the conductivity tensor we can relate  the potential $\phi _{\Kn}$ to the current induced in the system,
\begin{eqnarray}
{\bf J} _{\Kn} &=& {\underline{\sigma }} \,  {\bf E}_{\Kn}
 =  \frac {e  n_0}{\omega m^*} \Kn \phi _{\Kn} + \frac {e}{\omega m^*}
\sum _{\Kn '} \Kn ' \delta n _{\Kn-\Kn'} \phi _{\Kn '} \nonumber \\
&& - i \sigma _H (\hat z \times \Kn) \phi _{\Kn} - \frac i 2 \frac {e  \hbar} {\delta m ^*}
n_0 (\hat z \times \Kn ) \phi _{\Kn} \nonumber \\
&&-\frac i 2 \frac {e  \hbar }{\delta  m^*} \sum _{\Kn '} ( \hat z \times \Kn ') \delta n _{\Kn-\Kn '}
\phi _{\Kn '},
\end{eqnarray}
where ${\bf E}_{\Kn}$=$i \Kn \phi _{\Kn}$ is Fourier component of the plasmon electric field.
${\bf J} _{\Kn}$ can be related to the density $\delta  \rho _{\Kn}$ using the continuity equation, and finally we introduce self-consistency by linking
the electric potential and the density of charge induced by the plasmon through the
Poisson equation,
$ \phi _{\Kn}$=$\frac {2 \pi e }{\epsilon _0 K } \, \delta  \rho _{\Kn}$.  Through these steps
we obtain for the plasmon potential
\begin{equation}
\left ( 1 - \frac {\omega _K ^2 }{ \omega ^2} \right ) \phi _{\Kn} = V (K) \sum _{\Kn '}
\Delta (\Kn,\Kn' ; \omega) \phi _{\Kn '}, \, {\rm with}
\label{scat1}
\end{equation}
\begin{equation}
\! \! \Delta (\Kn,\Kn '; \omega)  \! \! = \!  \! \frac 1 {m^*}  \! \! \left [
\frac 1 {\omega ^2} \!  \Kn \cdot  \!  \Kn '\!  - \!  \frac i {\omega}  \frac \hbar {2 \delta}
\hat z \! \cdot  \! (\Kn \!  \! \times \! \!\Kn ') \right ] \delta n _{\Kn \! - \! \Kn '} \,  .
\label{Delta}
\nonumber
\end{equation}
Equation \ref{scat1} describes scattering between plasmons of momentum $\Kn$ and $\Kn '$
due to a charge modulation $ \delta n _{\Kn -\Kn '}$.
The scattering has two terms, a real symmetric term proportional to $\Kn \cdot  \Kn '$ that produces reciprocal scattering, and an asymmetric imaginary  term proportional to $\Kn \times  \Kn '$ which  induces skew trajectories for scattered plasmons. Note that this last term is proportional to the  Berry's curvature of the one-electron band structure, which enters through the Hall conductivity $\sigma _{xy}$.  For high frequency plasmons, where one does not expect quantum structure to be apparent, skew scattering vanishes \cite{SI}.

The impurity in this approach is implicitly present through its impact on the density.  The induced density is given approximately by
\cite{Ando:1982aa,Vignale:book,note2}
\begin{equation}
\delta n_{K} = \chi_0 {V_{imp}(K)}/{\epsilon (K)}
\end{equation}
where $\epsilon (K)$ = $1+ \frac {q_{TF} } K$ and $\chi _0$ =$-\frac {m^*}{ 2 \pi \hbar ^2} $ are  the static dielectric constant of the uniform  electron gas and
the static density-density response function  respectively. Here $q_{TF}$ =$
\frac { m^* e^2}{\hbar ^2 \epsilon _0}$ is the Thomas-Fermi wavevector of the electron gas.
In the case where the
charge modulation is created by
a Coulomb  impurity of charge $Ze$, and for long wavelengths,
the density modulation that appears in Eq. \ref{Delta} is independent of the wavevector,  $\delta n _{K}$=$Ze$.

An alternative derivation of Eqs. \ref{scat1} and \ref{Delta} can be carried out using an RPA dielectric formalism that is perturbative in the screened impurity potential, for which one does {\it not} find skew scattering when the QGD vanishes \cite{Sziklas:1965aa,Tripathy:1969aa,Rudin:1993aa,Rudin:1996aa,Torre:2017aa}.
By contrast, we find in applying this formalism to systems with non-vanishing QGD, skew scattering is indeed present, with precisely the form found in the macroscopic analysis above (see
SI \cite{SI}.)


\vspace{0.1cm}
\par \noindent {\it Scattering Cross Section.}
We have shown that an impurity of charge $Ze$ acts as a scattering center for plasmons. To obtain its scattering cross section, we rewrite Eq.\ref{scat1} in the form of a Lippman-Schwinger equation,\cite{Sziklas:1965aa,Torre:2017aa}
\begin{equation}
\phi_{\Kn}=\phi ^0_{\Kn}+\frac {\omega ^2 V(K)}{\omega ^2 -\omega _K^2 + i \eta}\sum _{\Kn '} \Delta (\Kn,\Kn' ; \omega) \phi_{\Kn '},
\label{LS}
\end{equation}
where $ \phi ^0_{\Kn}$ is the solution of the homogeneous equation $(\omega ^2 -\omega _K ^2) \phi^0 _{\Kn}=0$.   Solving Eq.\ref{LS} involves
boundary conditions,  for which we assume that $ \phi ^0_{\Kn}$  corresponds to an incident plane wave of momentum $\Kn$ and frequency $\omega$.  In the first order Born approximation
one finds
\begin{eqnarray}
& & \phi_{\Kn} (r,\theta) \,
\xrightarrow
    {{\rm r} \rightarrow \infty}
\, e ^{i \Kn \rn}- f(\theta,\omega) \frac {e ^{i K r}}{\sqrt{r}}, \, \, \, \, \, \,  {\rm with} \, \,
\nonumber \\
f(\theta,\omega) &  \! = \!  & \frac { e^{i \pi /4} K^{3/2} \sqrt{2 \pi} }{E_F} \frac { e^2 Z}{\epsilon _0 } \frac 1 {2K|\sin \frac {\theta} 2| + q_{TF}} \times  \nonumber \\
& & \! \left ( \cos \theta  \! - \! i  \frac {\hbar \omega }{2 \delta}\sin{\theta}  \! \right ).
\label{ampl}
\end{eqnarray}
Here $f(\theta,\omega)$ is the scattering amplitude, which depends both on the plasmon frequency and on the angle, $\theta$, formed by the incident and the scattered plasmons.
This expression demonstrates that an asymmetry appears in the electric potential energy fluctuations created by plasmons scattering from a point-like Coulomb impurity.  That this  occurs is {\it only} possible because the underlying probability amplitude for scattering of a plasmon is itself asymmetric, a signal that the plasmon carries a non-vanishing QGD.  An observation of non-reciprocity in the electric field of scattered plasmons thus signals the non-trivial quantum geometry of their Hilbert space.

 \begin{figure}[tbp]
\includegraphics[width=8.5cm,clip]{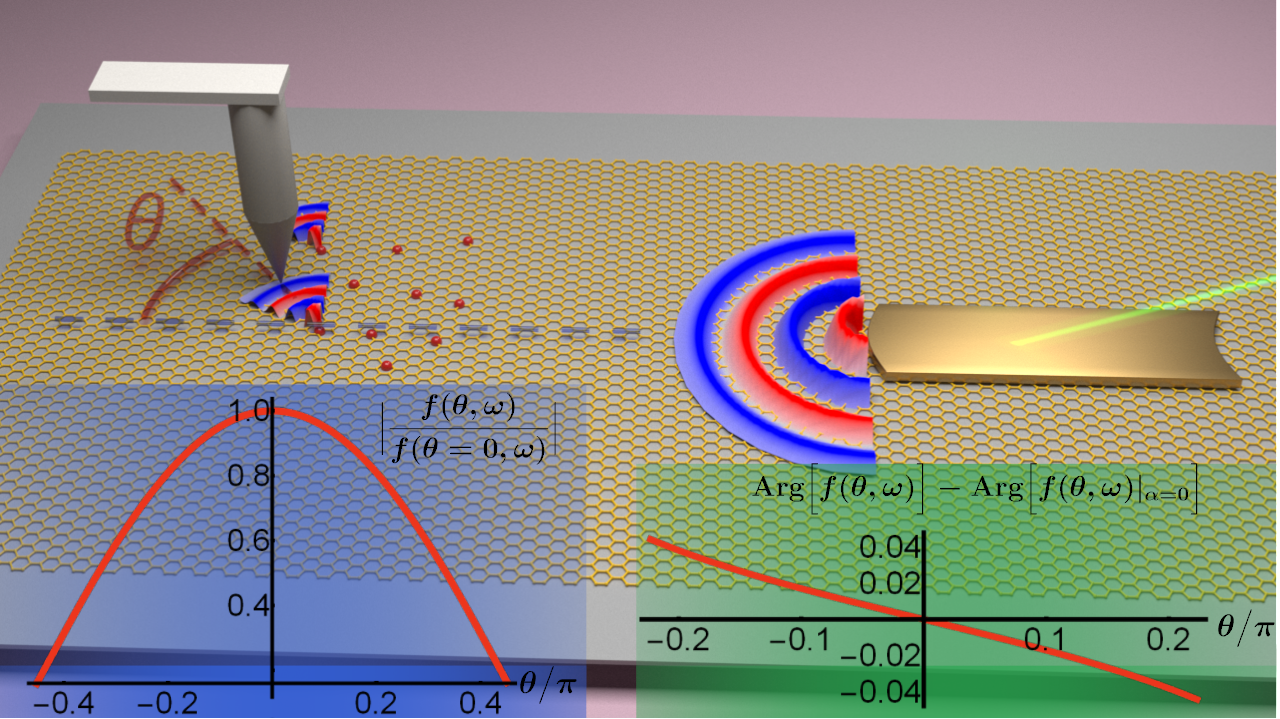}
\caption{Main figure:
Proposed geometry to detect skew scattering due to QGD carried by a plasmon.  Light impinging upon an antenna coupled to a two dimensional material launches plasmons.  A tip placed near an impurity detects the plasmon scattering amplitude as a function of angle $\theta$.  Because the tip collects information about both the magnitude and phase of the scattered wave, skew scattering asymmetry will be evident.  Left inset: Magnitude of scattered contribution to electric potential $f(\theta,\omega)$ (Eq. 17) normalized to $\theta=0$, as a function of scattering angle $\theta$ for $\alpha = 0.854 {\AA}^{2}$.  Right inset: Difference in phase angle for systems with $\alpha=0.854 {\AA}^2$ and $\alpha=0$.  Other parameters used for insets: $\omega=0.2$eV, $\delta=1.9$eV, $E_F-\delta=0.04$eV, $q_{TF}=27.87{\AA}^{-1}$.  $K$ is fixed by the plasmon dispersion relation.
}
\label{Figure2}
\end{figure}

\vspace{0.1cm}
\par \noindent{\it Discussion.}
In this work we demonstrated
that plasmon wavefunctions may support non-trivial internal structure, specifically a dipole moment, tied to their quantum geometry.  While quantum geometric effects are well-known for single-particle properties of some materials \cite{Resta_1994,Chang:1995aa,Calderon:2001aa,Fang_2003,Haldane_2004,parkin:2008aa,Beenakker:2009aa,
Xiao_2010,Hasan_2010,Gradhand_2012,Fert:2013aa,Nagaosa:2013aa,
Bliokh:2015aa,Vanderbilt_book,Brey:2017aa,Armitage_2018}, their impact on collective excitations -- particularly excitons -- are only more recently appreciated
\cite{Yao_2008,Kuga_2008,Garate_2011,Srivastava_2015,Zhou_2015,Qiu:2015aa,Trushin_2016,Wu_2017,
Trushin_2018,Hichri_2019,Kwan_2020,Cao_2021}.  Our study demonstrates for the first time that such quantum effects are also relevant for plasmons.

The QGD appears for materials where the underlying band structures carry non-trivial Berry's curvature. While this is the case, for example, for ultrathin TMDs, these semiconductors have degenerate Dirac-like
gaps at different points of the Brillouin zone, such that when doped the
Berry's curvature effects from different valleys will cancel.
However, in monolayer MoS$_2$ it
is possible to imbalance populations of carriers in these valleys
by optical pumping with circularly polarized light \cite{Mak:2012aa,Zeng:2012aa}, providing a route to lifting precise cancellation of asymmetries by the valleys.  Another possibility involves applying a magnetic field or doping with magnetic ions \cite{Fu_2020}. Because of the locking of spin and valley indices in these materials \cite{Xiao_2012}, magnetization in the system induces differing Fermi wavevectors in the valleys, eliminating precise cancellations in their plasmon dynamics.

The asymmetric skew scattering of plasmons by charged impurities we propose in this paper can be observed experimentally using different experimental setups. One possibility is to send a plasmon wave-packet from  a  scanning near field tip \cite{Chen:2012aa,Fei:2012aa} or a fixed nano-antenna \cite{Alonso-Gonzalez:2014aa} towards a set of charged impurities, {or a large defect that creates a cylindrical charge density modulation}, and analyze, as a function of the angle, the electric field of the reflected plasmons by their coupling to another scanning near field tip. (See Fig. \ref{Figure2}.)   Note that
the ratio of the symmetric and antisymmetric components of the scattering cross section for the electric potential (Eq. \ref{ampl})
is of order the ratio between the plasmon energy and the gap of the host semiconductor. In the case of MoS$_2$ the latter is roughly 1.9$eV$ \cite{Mak:2010ab}, while plasmon energies are in the range of 10-100$meV$. Thus we expect the skew scattering to be of order  1-5$\%$ of the symmetric one, which should be measurable.

Non-reciprocal scattering from impurities presents a new way to guide two-dimensional plasmons, an ability of great interest for
technological applications.  More generally, non-reciprocity in plasmon dynamics is of considerable fundamental interest, both in homogeneous systems \cite{Shi_2018,Duppen:2016aa,Papaj:2020aa,Morgado:2018aa}
and in more structure environments
\cite{Song:2016aa,Kumar:2016aa,Principi:2016aa,Jin:2016aa,Brey:2020aa}.
Although plasmons are typically described by macroscopic electric fields and currents,
in this work we have shown that they can support behaviors rooted in microscopic wavefunctions and their quantum geometry.  In the case of plasmon skew scattering, the non-reciprocity is tied to a measure of this which is specifically two-body in nature, the quantum geometric dipole.  Detecting such scattering would give a first view on the microscopic, internal structure of these fundamental excitations of metallic systems.

{\it Acknowledgments.} L.B. acknowledges funding from PGC2018-097018-B-I00 (MICIN/AEI/FEDER, EU).
HAF and JC acknowledge the support of the NSF  through Grant Nos. DMR-1914451 and ECCS-1936406.
HAF acknowledges support from the US-Israel
Binational Science Foundation (Grant Nos. 2016130 and 2018726),
of the Research Corporation for Science Advancement through a Cottrell SEED Award.


\begin{thebibliography}{83}
\bibitem [{not()}]{note2}%
  \BibitemOpen
  \href@noop {} {}\bibinfo {note} {Note that in this work all the response
  functions corresponds to an spinless two-dimensional electron
  gas.}\BibitemShut {Stop}%
\end{thebibliography}
%

\end{document}